\documentclass[prl,twocolumn,amsmath,amssymb,groupedaddress,superscriptaddress]{revtex4}
\usepackage{graphicx}

\def\be{\begin{equation}}
  \def\ee{\end{equation}}

\begin{document}
\title{Universal algebraic growth of entanglement entropy \\ in many-body localized systems with power-law interactions}

\author{Xiaolong \surname{Deng}}
\email{Xiaolong.Deng@itp.uni-hannover.de}
\affiliation{Institut f\"ur Theoretische Physik, Leibniz Universit\"at Hannover, Appelstr. 2, 30167 Hannover, Germany}
\author{Guido Masella}
\affiliation{ISIS (UMR 7006) and icFRC, University of Strasbourg and CNRS, 67000 Strasbourg, France}
\author{Guido Pupillo}
\affiliation{ISIS (UMR 7006) and icFRC, University of Strasbourg and CNRS, 67000 Strasbourg, France}
\author{Luis Santos}
\affiliation{Institut f\"ur Theoretische Physik, Leibniz Universit\"at Hannover, Appelstr. 2, 30167 Hannover, Germany}

\begin{abstract}
Power-law interactions play a key role in a large variety of physical systems. In the presence of disorder, these systems
may undergo many-body localization for a sufficiently large disorder. Within the many-body localized phase the system
presents in time an algebraic growth of entanglement entropy, $S_{vN}(t)\propto t^{\gamma}$. Whereas the critical disorder for many-body localization depends on the
system parameters, we find by extensive numerical calculations that the exponent $\gamma$ acquires a universal value $\gamma_c\simeq 0.33$ at the many-body localization transition,
for different lattice models, decay powers, filling factors or initial conditions. Moreover, our results suggest an intriguing relation between $\gamma_c$ and the critical minimal decay power of interactions 
necessary for many-body localization. 
\end{abstract}
\date{\today}
\maketitle




The interplay between disorder and interactions plays a key role in the understanding of transport and thermalization in 
many-body quantum systems. Whereas quantum interference leads to Anderson localization in non-interacting disordered systems~\cite{Anderson1958},  
localization may occur even in highly excited states in the presence of interactions~\cite{Basko2006}. 
Many-body localization~(MBL) is of fundamental relevance in quantum statistical mechanics, being the only known robust mechanism that may prevent thermalization in 
an isolated system. As a result, MBL has attracted a huge attention in recents years~\cite{Nandshikore2015,Altman2015,Abanin2017,Alet2018,Abanin2019}, including 
breakthrough experiments~\cite{Schreiber2015,Choi2016,Bordia2016,Smith2016,Lueschen2017,Lukin2019,Rispoli2019}.


Whereas MBL research has mostly focused on local interactions, recent works are unveiling the intriguing  
thermalization and MBL physics in disordered systems with power-law interactions~\cite{Burin2006,Pino2014,Yao2014,Burin2015a,Burin2015b,Hauke2015,Li2016,Gutman2016,Singh2017,Nandkishore2017,
Tikhonov2018,Safavi2019,DeTomasi2019,Nag2019,Roy2019,Botzung2019,Choi2019,Maksymov2019,Schiffer2019}. On one hand, this is justified by the possible relevance of long-range interacting systems for the understanding 
of MBL in dimensions larger than one~\cite{Singh2017,Kloss2019}. On the other hand, power-law interactions~(van der Waals, dipolar, Coulomb, or even of variable power) 
are fundamentally relevant for a large variety of physical systems, including nuclear spins~\cite{Alvarez2016}, nitrogen vacancy centers in diamonds~\cite{Waldherr2014}, 
polar molecules~\cite{Yan2013}, magnetic atoms~\cite{DePaz2013, Baier2016}, 
Rydberg gases~\cite{Zeiher2017,Leseleuc2019}, atoms at photonic crystals~\cite{Hung2016}, and trapped ions~\cite{Richerme2014,Jurcevic2014}. 
MBL is expected for a sufficiently quickly decaying power-law interactions~(up to the delocalizing effect of rare ergodic spots~\cite{DeRoeck2017}).
One-dimensional XXZ spin models, with both Ising and exchange interactions decaying with the interparticle distance $r$ as $1/r^a$, have been predicted to present 
MBL for a sufficiently large disorder for $a>a_c=2$~\cite{Burin2006,Yao2014,Burin2015a}. 
For XY models, with just spin exchange, MBL 
has been predicted for $a>a_c=3/2$ due to emerging Ising interactions~\cite{Burin2015b}, although recent numerical calculations indicate that $a_c$ may be 
smaller~\cite{Safavi2019,Roy2019,Schiffer2019}.



\begin{figure*}[t]
  \begin{center}
  \includegraphics[width=2\columnwidth]{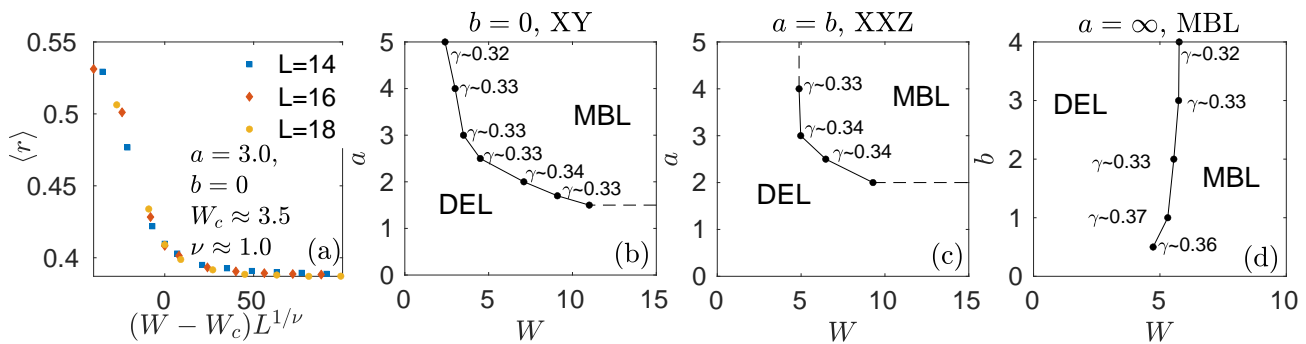}
  \caption{(a) $\langle r \rangle$ for $b=0$, $a=3$ as function of $W$ for $L=14$, $16$, and $18$. 
  Phase diagrams, evaluated using the level spacing statistics, as a function of $W$ and the decay power $a$ for (b) XY model~($b=0$), (c) XXZ model~($a=b$ and $V=1$), and 
  (d) EHM model~($a=\infty$, $V=1$). We indicate the value of the power $\gamma$ of the growth of $S_{vN}(t)$ at the MBL transition.}
  \vspace{-0.5cm}
  \label{fig:1}
  \end{center}
\end{figure*}



Entanglement dynamics within the MBL phase presents intriguing features. In particular, whereas the entanglement entropy, $S_{vN}$,
saturates in the Anderson localized case to a system-size independent value, in MBL systems 
entanglement propagates due to dephasing even in the absence of energy or particle transport. In particular, for local 
interactions, $S_{vN}$ grows logarithmically in time until reaching a volume-law value~\cite{Znidaric2008,Bardarson2012,Serbyn2013,Vosk2013}. This entropy 
growth results from the interaction between exponentially localized local integrals of motion~(LIOMs)~\cite{Serbyn2013b,Huse2014} adiabatically connected with the 
single-particle states. In contrast, in power-law interacting systems single-particle localization is rather algebraic~\cite{Deng2016,Deng2018}, and a similar LIOM logic predicts algebraic entropy growth~\cite{DeTomasi2019}, as observed numerically~\cite{Pino2014,Safavi2019,DeTomasi2019}.


In this Letter, we investigate the MBL phase of hard-core bosons with power-law hops and interactions, or equivalently spin models with power-law exchange and Ising terms. 
Using exact diagonalization, we determine the onset of MBL from level spacing statistics 
for different models of experimental relevance: XY model, XXZ model with equal decay power for 
Ising and exchange interactions, and extended-Hubbard model~(EHM) with nearest-neighbor~(NN) hops and power-law interactions. 
Using exact evolution and Krylov techniques, we analyze the entanglement dynamics, and in particular 
 the algebraic growth $S_{vN}(t)\propto t^\gamma$. In contrast to previous studies, which concentrated in a narrow range of $a$ values and 
 a single disorder strength $W$ well within the MBL regime~\cite{Safavi2019,DeTomasi2019}, we analyze in detail the dependence on $W$, showing 
that the algebraic growth presents a remarkable universality at the critical disorder $W_c(a)$ that marks the onset of MBL. At criticality, 
$\gamma=\gamma_c\simeq 0.33$, for the XY, XXZ and EHM models, irrespective of the decay power of the interactions, the filling factor, or the initial state. 
Interestingly, our results suggest a surprising relation between 
$\gamma_c$ and the critical $a_c$ for MBL.



\paragraph{Model.--} We consider hard-core bosons in a disordered 1D lattice, which present both 
power-law hopping and interactions. The system is described by the Hamiltonian:
\begin{eqnarray}
\hat H = &-&J\sum_{i,j\neq i} \frac{1}{|r_i-r_j|^a} (\hat b_i^\dag\hat b_j + \mathrm{h.c.}) \nonumber \\
&+&V \sum_{i,j\neq i}\frac{1}{|r_i-r_j|^b}\hat n_i \hat n_j + \sum_j\epsilon_j \hat n_j,
\label{eq:H}
\end{eqnarray}
where $\hat b_j$ are bosonic operators at site $j$~($(\hat b_j^\dag)^2=0$), $\hat n_j=\hat b_j \hat b_j$,  
$J=1$ and $V$ are, respectively, the hopping amplitude and interaction strength to nearest-neighbors~(NN), and $\epsilon_j$ is a random on-site energy uniformly distributed between $-W$ and $W$.
Hamiltonian~\eqref{eq:H} is interesting for a large variety of physical problems. It may be mapped to a spin-$1/2$ Hamiltonian with power-law exchange and Ising terms, and random on-site magnetic field. 
In particular, when $V=0$~(or equivalently $b=0$, due to 
number conservation), Eq.~\eqref{eq:H} reduces to an XY Hamiltonian, as that already realized, in absence of disorder, in polar molecules with two available rotational states~\cite{Yan2013}, 
Rydberg atoms~\cite{Leseleuc2019}, and trapped ions~\cite{Richerme2014,Jurcevic2014}. 
The case $a=b$, which reduces to a power-law XXZ model, is directly relevant for two-component magnetic atoms~\cite{DePaz2013} and polar molecules in the presence of an external electric field. 
Finally, for $a=\infty$ the model reduces to the extended Hubbard model~(EHM) with NN hopping and power-law interactions, already realized in magnetic atoms polarized in the maximally stretched state~\cite{Baier2016}.
Below, we focus on the localization properties and dynamics of these three experimentally relevant models.



\paragraph{Many-body localization.--}  We first establish, for the different cases, the critical disorder strength to achieve MBL. 
We consider a lattice of $L$ sites with open-boundary conditions. By means of exact diagonalization for up to $L=18$ sites, we determine the eigenstates for $L/2$ hard-core bosons 
(although we focus below on half-filling, we have found similar results for other filling factors~\cite{footnote:SM}). 
We study the level-spacing statistics, characterized by $r_n = \min(\delta_n,\delta_{n-1})/\max(\delta_n,\delta_{n-1})$, where $\delta_n \equiv E_{n+1}-E_{n}$, and $E_n$ denotes the 
eigen-energies in growing order. We obtain $\langle r \rangle$ after averaging $r_n$ over all states with $|E_n|<W$~(in order to avoid spurious effects given by states at the spectral edges) 
and over up to $1000$ disorder realizations~(for the EHM the center of the averaging window is displaced to the maximum of the density of states~\cite{footnote:SM}). 
We determine $\langle r \rangle$ for different system sizes $L=14$, $16$, and $18$. 
In the thermodynamic limit, it is expected that integrable or MBL systems are characterized by a Poissonian level spacing distribution, characterized by $\langle r \rangle\simeq 0.386$, whereas 
ergodic systems present a Wigner-Dyson distribution in the Gaussian Orthogonal Ensemble, which results in $\langle r \rangle\simeq 0.529$ in the delocalized phase (DEL). 
The critical disorder strength $W_c$ marking the onset of MBL is then given by the crossing point of the $\langle r\rangle$ curves for different $L$, which is hence stationary under scaling of the system size.
In order to determine properly $W_c$, we perform a finite-size scaling analysis, expressing $\langle r \rangle$ as a function of $(W-W_c)L^{1/\nu}$, such that 
curves for different $L$ collapse, as illustrated in Fig.~\ref{fig:1}(a) for $b=0$ and $a=3$.



\begin{figure}[t]
  \includegraphics[width=1\columnwidth]{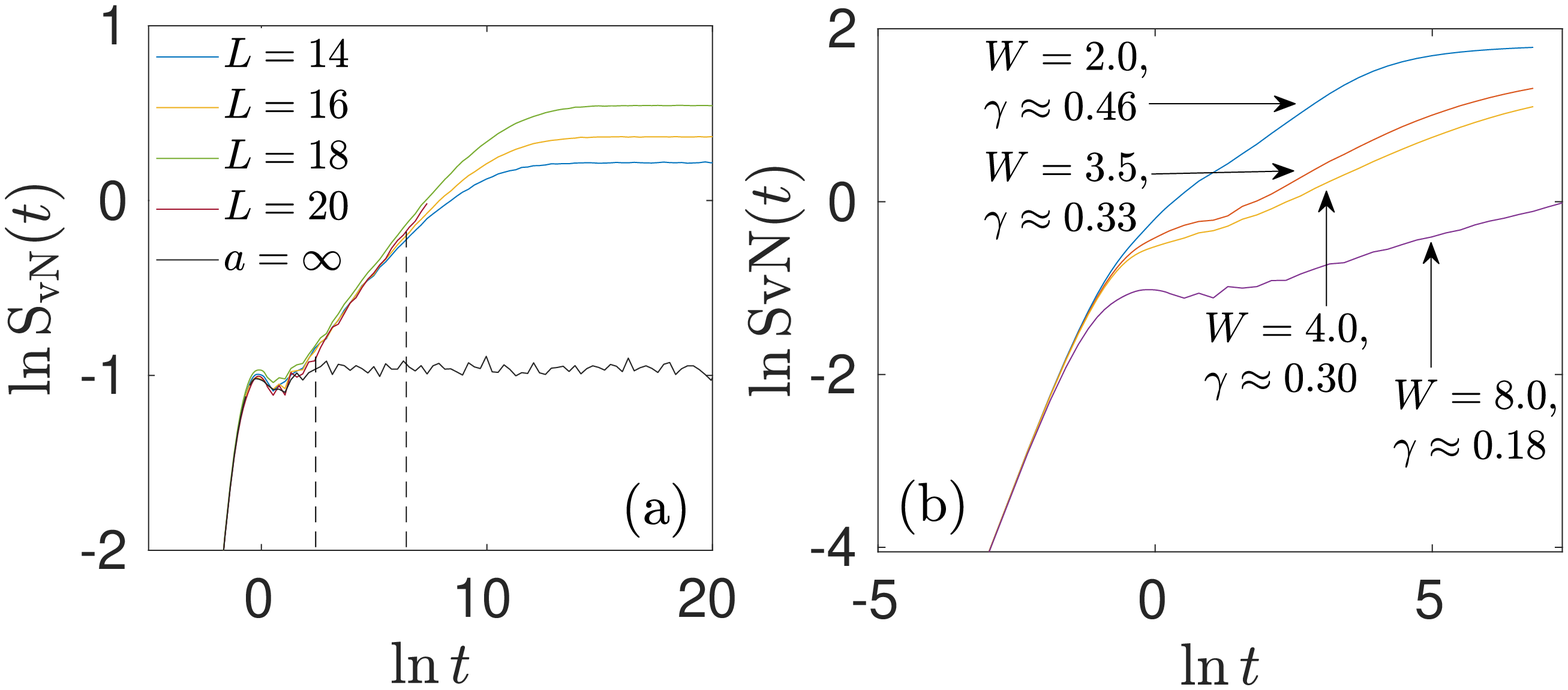}
  \caption{(a) Entanglement entropy $S_{vN}(t)$ for $a=3$, $b=0$, $W=8$ and different $L$. For $L=14$, $16$ and $18$ we employ exact evolution and $2000$ realizations, whereas for 
  $L=20$ we use the Krylov subspace method and $1000$ realizations. Note the relatively large time window~(between the dashed vertical lines) with a common algebraic growth for all $L$ values that we employ 
  for fitting the value of the power $\gamma$. We also depict the curve for $a=\infty$~(Anderson localization). (b) $S_{vN}(t)$ for $a=3$, $b=0$, $L=20$. Larger $W$ results 
  in lower $\gamma$. For these values $W_c\simeq 3.5$.}
  \label{fig:2}
\end{figure}

\paragraph{Phase diagrams.--} Figure~\ref{fig:1}(b) depicts the phase diagram as a function of $W/J$ and $a$ for the XY model~($b=0$ or $V=0$). For $a<a_c\simeq 3/2$, we observe, 
for the available system sizes in our numerics, no clear crossing point in the finite-size scaling of $\langle r \rangle$, and hence no trace of MBL. This is in agreement with the predictions of Ref.~\cite{Burin2015b}, 
although we cannot rule out that the critical power may be slightly lower~\cite{Safavi2019,Roy2019,Schiffer2019}.  
Figure~\ref{fig:2}(c) shows the XXZ case~($a=b$) with $V=1$. In agreement with Ref.~\cite{Burin2006,Burin2015a} we observe no MBL for $a<a_c\simeq 2$. 
Finally, Fig.~\ref{fig:2}(d) depicts the EHM with NN hopping, for which MBL occurs at any $b$ value, being enhanced for $b<2$~\cite{Roy2019}.  



\paragraph{Entanglement entropy.--} We are particularly interested in the entanglement dynamics within the MBL region.  
In the following, we consider that the system is initially prepared in a half-filled density wave state $\dots 101010 \dots$. This choice is inspired by experiments~\cite{Schreiber2015}, but other choices 
of the initial state and the filling factor do not alter the results~\cite{footnote:SM}. 
The entanglement dynamics is monitored by means of the entanglement entropy, $S_{vN}(t)=-\mathrm{Tr}[ \hat \rho_A \mathrm{ln} \hat \rho_A ]$, with 
$\hat\rho_A=\mathrm{Tr}_B[\hat\rho]$ the reduced density matrix of the left half of the system~($A$) when tracing over the other half~($B$). 

For systems up to $L=18$ sites, we determine the dynamics at any time $t>0$ using exact evolution. Krylov subspace techniques allow us  
calculations with larger system sizes~(up to $L=22$) but they are limited to moderate time scales. We have checked that for a given disorder realization and up to $L=18$ 
the Krylov and the exact calculation provide the same result within $10^{-6}$ relative error in the determination of $S_{vN}(t)$. A large number of disorder realizations is crucial
to achieve good statistics and converging results for $\gamma$,  since anomalous regions of small disorder increase the value of $\gamma$. 
For $L=14,16,18$ we choose up to $2000$ samples, and for $L=20$ up to $1000$ samples, which lead to converging results~\cite{footnote-DMRG}.



Figure~\ref{fig:2}(a) shows our results for the XY model~($b=0$) with $a=3$ and $W=8$, which illustrate a typical entropy growth in our calculations. Initially, 
$S_{vN}(t=0)=0$ since we start with a Fock state. At short times local dynamics leads to an entropy growth that is independent of the power $a$, and indeed it is shared by the $a=\infty$~(NN) case, which presents Anderson localization. 
After this initial dynamics, for finite $a$, the entropy grows algebraically, $S_{vN}(t)\propto t^\gamma$, until saturating at a value that depends on the system size. 
We note that the onset of the algebraic growth is delayed to longer times when $a$ grows, resulting in an entropy plateau shared with the NN case.
The onset time diverges when $a\to\infty$. 
As shown in Fig.~\ref{fig:2}(a), the slope of the log-log curves converges within various decades for $L=14$, $16$, $18$ and $20$~(the latter obtained with the Krylov method, and hence limited to shorter times), 
allowing us to exclude finite-size effects in the determination of $\gamma$ by fitting only within the converged time window.  



\begin{figure} [t]
  \includegraphics[width=\columnwidth]{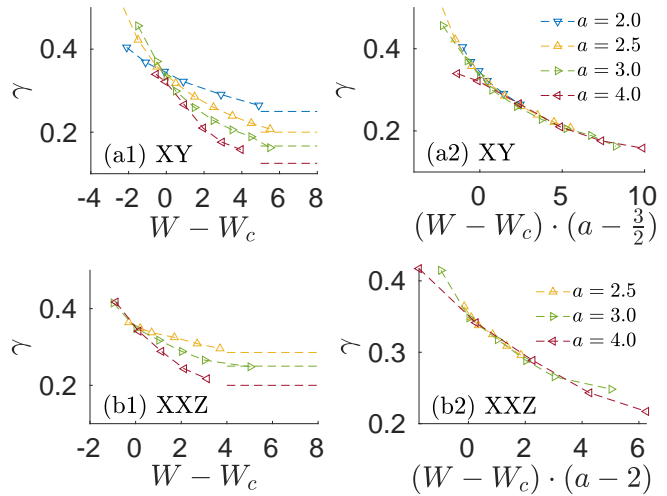}
  \caption{(a1) and (b1) Exponent $\gamma$ as a function of $W-W_c(a)$ for the XY model~($b=0$) and the XXZ model~($a=b$, $V=1$) for different $a$ values. 
  Note that all curves cut at $W=W_c$ indicating universality of $\gamma$ at the onset of MBL. The horizontal lines indicate $1/2a$~(XY model) and $1/(a+1)$~(XXZ model).
  (a2) and (b2) Same figure, but as a function of $(W-W_c(a))(a-3/2)$ for the XY model, and of $(W-W_c(a))(a-2)$ for the XXZ mode. Note that the curves collapse in the vicinity of the onset of MBL, 
  indicating a universal scaling. }
  \label{fig:3}
\end{figure}




\paragraph{Universal entropy growth in the XY model.--} The value of the exponent $\gamma$ depends on both the disorder strength $W$ and the powers $a$ and $b$. 
Figure~\ref{fig:2}(b) depicts our results for the XY model with $a=3$ and $L=20$ sites, for $W$ 
around the critical $W_c\simeq 3.5$. Both for the MBL and the extended phase we have an algebraic growth. Although our focus is on the MBL regime, 
we note that the entropy growth in the extended regime is far from linear as one would expect for an ergodic system~\cite{Kim2013}, 
indicating possible non-ergodicity of the extended regime, at least in the vicinity of the MBL phase.
The power $\gamma$ decreases with growing $W$, with a critical power $\gamma_c\simeq 0.33$ at $W=W_c$. As we discuss in the following, this critical entropy growth turns out to be universal. 
This universality constitutes the main result of this paper. 

In Fig.~\ref{fig:3}(a1) we depict $\gamma$ for different $a$ values. 
Although the level-spacing analysis, which provides $W_c(a)$, and the study of the dynamics, which provides $\gamma$, are independent from each other, 
we observe that all $\gamma(a,W)$ curves converge at criticality, $W=W_c(a)$, at a value that within our numerical accuracy is approximately $\gamma_c\simeq 0.33\pm 0.02$~\cite{footnote-errorbars}. 
Hence, remarkably, the critical algebraic entropy growth at the MBL on-set is independent of $a$~(see also Fig.~\ref{fig:1}(b)). 
Moreover, for $W>W_c(a)$ in the vicinity of the MBL boundary,  $\gamma$ becomes to a good approximation a universal function of $(a-3/2)(W-W_c(a))$~(Fig.~\ref{fig:3}(a2)). 

For large $W$, we may expect that the LIOMs can be approximated by the population of single-particle states, which remain algebraically localized at lattice sites with the same power $a$ of the XY exchange~\cite{Deng2016,Deng2018}. 
Hence, the interaction between LIOMs placed at a distance $r$~(resulting from the hard-core constraint) should decay as $1/r^{2a}$. As a result, we would expect for large $W$, $\gamma=\gamma_\infty(a) =1/2a$ at large disorder. 
As shown in Fig.~\ref{fig:3}(a1), this is approximately the case~(dashed lines indicate $\gamma_\infty(a)$). 
Calculations with large system sizes would be however necessary to establish the asymptotic $\gamma_\infty(a)$ dependence more precisely, since 
due to the low saturation entropy we cannot perform reliable fits of $\gamma$ for $W>12$ for the system sizes we can evaluate.
Interestingly, if $\gamma=\gamma_c$ holds for the MBL transition all the way till $a=a_c$, and since $W_c(a_c)$ diverges, 
then we would expect $\gamma_c=\gamma_\infty(a_c)$. Note that this is indeed fulfilled for $\gamma_\infty(a)\simeq 1/2a$, $a_c\simeq 3/2$ and $\gamma_c\simeq 1/3$.



\paragraph{Universal entropy growth in other models.--} Interestingly, a similar analysis for the XXZ model with $a=b$
reveals that the algebraic growth of $S_{vN}(t)$ is also universal at the onset of MBL with the same exponent 
$\gamma_c\simeq 0.33$~(see Fig.~\ref{fig:3}(b1)). Moreover, within the MBL in the vicinity of $W_c(a)$ $\gamma$ is a universal function of $(W-W_c(a))(a-2)$~(see Fig.~\ref{fig:3}(b2)).
For large $W$, following the arguments of Ref.~\cite{Pino2014}, we may expect a dependence $\gamma_\infty(a)\simeq 1/(a+1)$. 
Our numerical calculations, which as for the XY model are limited to $W<12$, suggest that this is approximately the case. 
Similar as above, we would expect $\gamma_c=\gamma_\infty(a_c)$. Note that the latter would be fulfilled for  $\gamma_\infty(a)\simeq 1/(a+1)$, $a_c\simeq 2$, and $\gamma_c\simeq 1/3$.
The results for the XXZ and XY models hence suggest that there is an intriguing relation between the universal growth of $S_{vN}(t)$ at the MBL transition, given by $\gamma_c$, and 
the critical power $a_c$ for observation of MBL.

Finally, we have analyzed the dynamics in the EHM~($a=\infty$), see Fig.~\ref{fig:1}(d)~\cite{footnote-Pino},  obtaining as well a critical $\gamma_c\simeq 0.33$~(slight deviations 
for $b\geq 2$ can be attributed to the difficulty of determining reliably $W_c$ for small system sizes). Moreover, modified EHMs with random interaction and hopping signs also show the same 
universal entropy growth at the MBL transition~\cite{footnote:SM}.



\paragraph{Conclusions.--} Hard-core bosons with power-law-decaying hops and interactions in disordered 1D lattices, or equivalently spin models with power-law-decaying exchange and Ising terms 
present MBL for a sufficiently large disorder. By means of level spacing statistics, we have determined the MBL regime for three models of particular experimental relevance: XY model, XXZ model, 
and EHM model with power-law interactions. Due to algebraic localization of LIOMs, the entanglement dynamics is characterized by an algebraic growth of entanglement entropy. 
We have shown that for all models, power-law decays of the interaction terms, filling factors and initial conditions  
the algebraic entropy growth is characterized by a universal power, $S_{vN}(t)\propto t^{\gamma_c\simeq 1/3}$. 
This remarkable non-trivial universality was overlooked in previous studies, which focused 
on particular $W$ values for specific models and/or narrow windows of powers $a$~\cite{footnote-Safavi}. 
Moreover, our results suggest a relation between $\gamma_c$ and the critical power $a_c$ for the observation of MBL.
Interestingly, this relation may open the possibility to determine $a_c$ with experiments performed at any other $a>a_c$ power, in particular 
$a=3$ characteristic of systems with dipolar interactions. We expect that our work will trigger further theoretical work in determining the origin of such 
an universal growth. Moreover, our analysis opens interesting questions about the universality of other entanglement measures, as for example the Fisher information~\cite{DeTomasi2019,Safavi2019}, which 
may be more easily monitored experimentally. 


We thank A. Burin, H. Hu, J. Zakrzewski, A. Lazarides and S. Roy for interesting discussions. L.~S. and X.~D. acknowledge the support of the German Science Foundation~(DFG) (SA 1031/11, SFB 1227, and 
Excellence Cluster QuantumFrontiers). G.~M. and G.~P. were supported by the ANR 5 "ERA-NET QuantERA" - Projet "RouTe"~(ANR-18- QUAN-0005-01), and LabEx NIE. 
G.~P. acknowledges support from the Institut Universitaire de France (IUF) and USIAS. G.~M. was also supported by the French National Research Agency (ANR) through the
"Programme d'Investissement d'Avenir" under contract ANR-17-EURE-0024.



\newpage


\begin{thebibliography}{99}
\bibitem{footnote-errorbars} The error bar is determined by the accuracy of the linear fitting of the algebraic growth region in log-log plots such as Fig.~\ref{fig:2}(a). Error bars are typically 
smaller than the symbols depicted in Figs.~\ref{fig:3}.
\bibitem{footnote-Pino} Algebraic entanglement growth in this model was discussed in Ref.~\cite{Pino2014} for $V\ll 1$ and a fixed large disorder $W=8$ for up to $L=16$. We have checked that our 
$\gamma$ coincides with that of that reference for $L=16$, $b=1$ and $V=0.1$.
\bibitem{footnote-Safavi} We note that for the XY model, for the particular $W$ value and window of $a$ values discussed in Ref.~\cite{Safavi2019}, we obtain approximately the same 
exponent $\gamma$.
\end{thebibliography}
\end{document}


\title{Supplementary Material for "Universal algebraic growth of entanglement entropy \\ in many-body localized systems with power-law interactions"}

\author{Xiaolong \surname{Deng}}
\email{Xiaolong.Deng@itp.uni-hannover.de}
\affiliation{Institut f\"ur Theoretische Physik, Leibniz Universit\"at Hannover, Appelstr. 2, 30167 Hannover, Germany}
\author{Guido Masella}
\affiliation{ISIS (UMR 7006) and icFRC, University of Strasbourg and CNRS, 67000 Strasbourg, France}
\author{Guido Pupillo}
\affiliation{ISIS (UMR 7006) and icFRC, University of Strasbourg and CNRS, 67000 Strasbourg, France}
\author{Luis Santos}
\affiliation{Institut f\"ur Theoretische Physik, Leibniz Universit\"at Hannover, Appelstr. 2, 30167 Hannover, Germany}

\begin{abstract}
We provide further detailed information on the universality of entanglement growth at the many-body localization~(MBL) transition.
We first show that the universality is maintained when considering different initial conditions and filling factors. 
We discuss as well further details about the calculations for the special case of the extended Hubbard model. Finally, we show that the observed 
universal growth of entanglement entropy is valid as well for an extended Hubbard model with random signs.
\end{abstract}

\date{\today}
\maketitle



\section{Universality with respect to the initial condition}
\begin{figure}[t]
  \begin{center}
    \includegraphics[width =1.0\columnwidth]{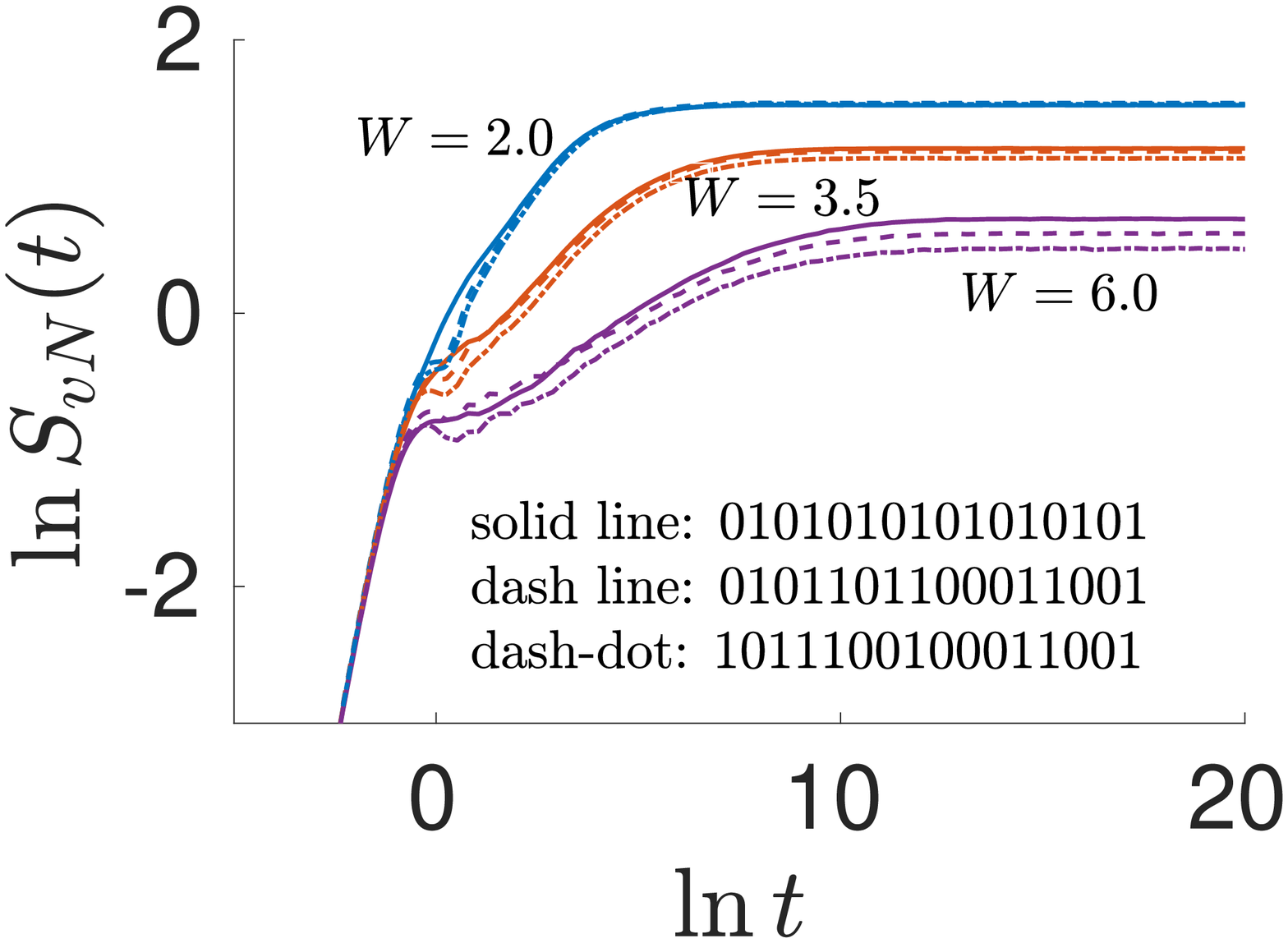}
    \vspace*{-0.6cm}
    \caption{(Color online) Evolution of the entanglement entropy of the XY model (with $a=3$) starting from randomly chosen initial states in the delocalized phase ($W=2.0$), critical phase ($W=3.5$) and many-body localized phase ($W=6.0$), respectively. The system size is $L=16$, and $1000$ disorder samples were considered for the statistics. Note that the saturation entropy may be slightly different, 
    but the slopes are the same for all initial states.}
    \label{fig:SM1}
    \vspace{-0.6cm}
  \end{center}
\end{figure}

In the main text, we discussed how the exponent $\gamma$ of the algebraic entanglement entropy growth depends on the disorder strength $W$, 
and the interaction powers $a$ and $b$. Although in the main text we discussed only the case of an initial density-wave state of the form 
$\dots 01010 \dots$, we have performed also numerical calculations starting from other initial states, randomly chosen from a general ensemble.
We observe no dependence on the initial condition chosen. As an illustration of this point, we show in Fig. \ref{fig:SM1}  
the evolution of the entanglement entropy for three different initial states for the case of the XY models with $a=3$.  
Although the saturation entropy slightly depends on the initial state, the slope of the growth in the log-log scale, and hence the value of $\gamma$ is independent of the 
initial state chosen. Hence, the observed universal algebraic growth of entanglement entropy at the MBL transition remains valid 
for a general ensemble of initial states.



\section{Universality with respect to different filling factors}
\begin{figure*}[t]
  \begin{center}
    \includegraphics[width =1.5\columnwidth]{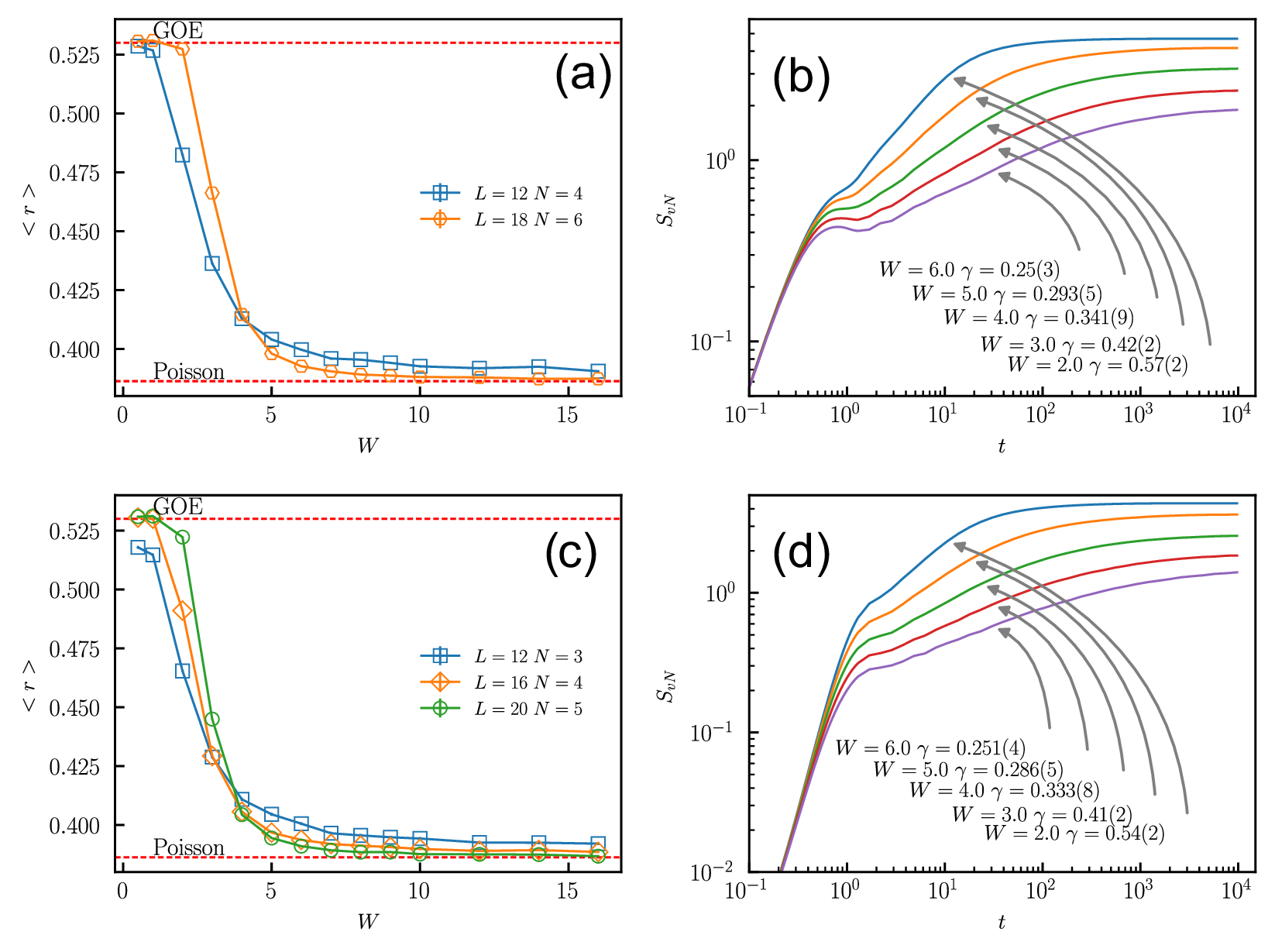}
    \vspace*{-0.6cm}
    \caption{(Color online) Results for filling factor $1/3$ (a, b) and $1/4$ (c, d) for the XXZ model with $a=b=3$.  
    Figures (a) and (c) depict the level statistics factor $\langle r \rangle$ as a function of the disorder strength $W$ for different system sizes $L$. 
    The crossing of the curves marks the transition, that occurs for $W_c = 4.1(1)$ for filling $1/3$, and for  $W_c = 3.8(1)$ for filling $1/4$. 
    Figure (b) and (d) show the evolution of the entanglement entropy as a function of 
    time for different values of $W$ ($L=18$ for filling $1/3$ and $L=20$ for filling $1/4$).}
    \label{fig:SM2}
    \vspace{-0.6cm}
  \end{center}
\end{figure*}

In the main text, we analyzed only the case of half-filling (or equivalently zero magnetization for spin systems). We have checked that the universality of the entanglement entropy 
growth at the MBL transition is maintained for other filling factors. 
Figure~\ref{fig:SM2} shows our results for fillings $1/3$ and $1/4$ for the XXZ model with $a=b=3$. In both cases we obtain $W_c\simeq 4$. At that value, we observe, as for the 
half-filling case, algebraically growing entanglement entropy growth with exponent $\gamma_c\simeq 0.33$. 



\section{Density of states of the EHM}
\begin{figure}[t]
  \begin{center}
    \includegraphics[width =1.0\columnwidth]{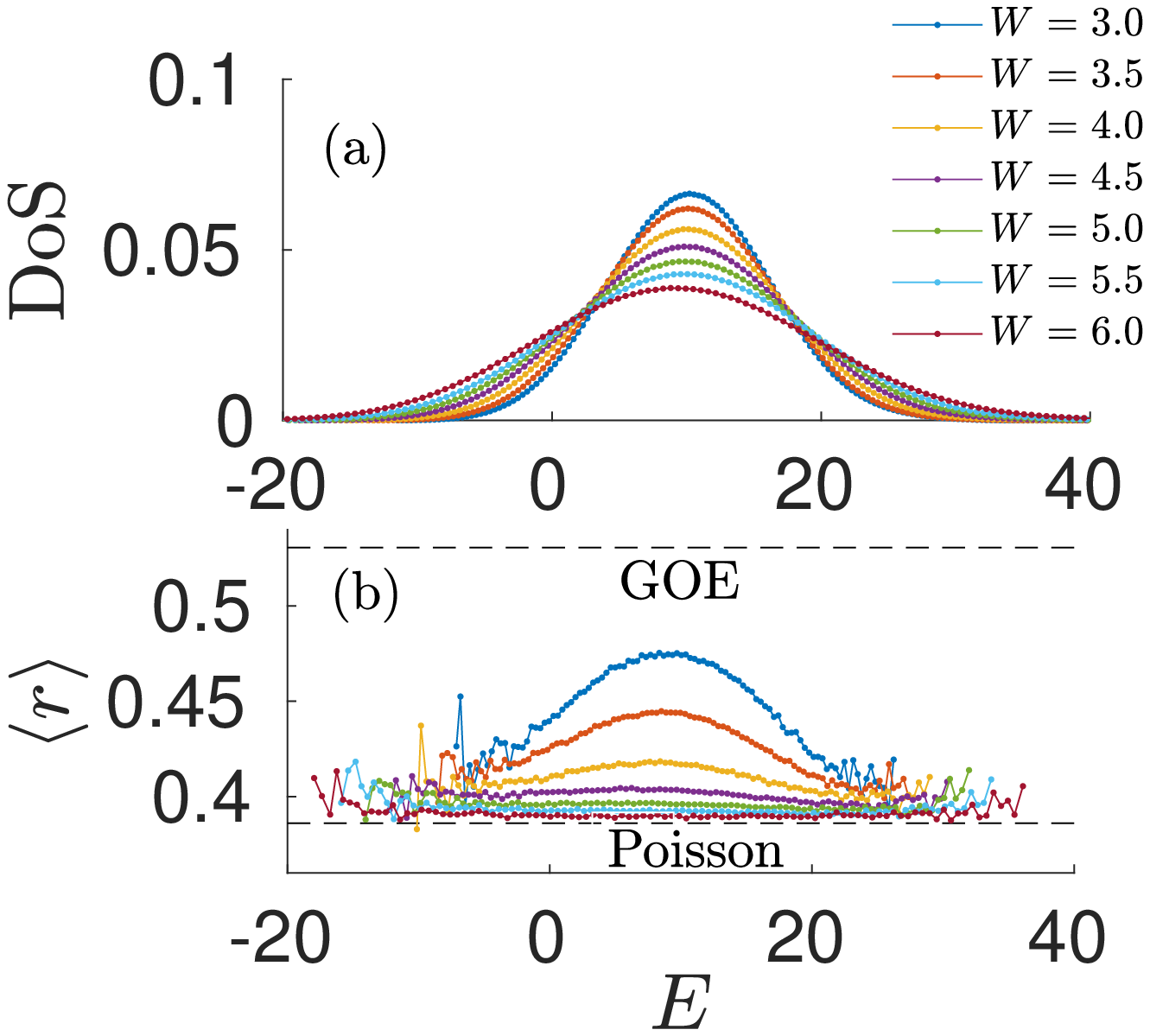}
    \vspace*{-0.6cm}
    \caption{(Color online)  (a) Density of states~(DoS) and factor $\langle r \rangle$  for different disorder strengths $W$  for the EHM with $b=1$. 
    We evaluate $1000$ samples for a system size $L=18$.}
    \label{fig:SM3}
    \vspace{-0.6cm}
  \end{center}
\end{figure}

When determining $W_c$ from level spacing statistics, we may choose to average $\langle{r}\rangle$ over all energies, which may be compromised by the spurious effect of states at the spectral edge, 
or rather average in the vicinity of the maximum of the density of states~(DoS). In the XY and XXZ models discussed in the main text, the maximum of the DoS is at energy $E=0$, justifying our 
average in a window $|E|<W$. In contrast, in the EHM the maximum of the density of states is at an energy $E_{\max(\mathrm{DoS})}\neq 0$, which is system dependent. In Fig.~\ref{fig:SM3}~(a)
we depict the DoS for $L=18$, which is peaked at $E_{\max(\mathrm{DoS})}=10$. The corresponding level-spacing factors $\langle r \rangle$ are depicted in Fig.~\ref{fig:SM3}~(b). Hence, 
in the case of the EHM, we average $\langle{r}\rangle$ rather over $|E-E_{\max(\mathrm{DoS})}|<W$. For Fig.~1(d) of the main text, we employed $8000$ disorder samples 
for $L=16$, and $1000$ ones for $L=18$. As shown in Fig.~1 of the main paper, $\gamma_c\simeq 0.33$ at $W_c$ also for the EHM model. We attribute the deviations observed for $b<2$ to the 
relatively imprecise determination of $W_c$, which for $b<2$ demands larger system sizes than those we can evaluate numerically. 



\section{Extended Hubbard model with random signs}

In the main text we considered an EHM with isotropic regular hopping and interacting terms. However, a more general EHM Hamiltonian has 
been recently theoretically considered in the context of MBL research~\cite{Yao2014,Gutman2016,Tikhonov2018}, which is of the form:
\begin{eqnarray}
\hat H &=& -J\sum_{\langle i,j\rangle } u_{ij} (\hat b_i^\dag\hat b_j + \mathrm{h.c.}) \nonumber \\ 
&+&  V \sum_{i,j\neq i}\frac{v_{ij}}{|r_i-r_j|^b}\hat n_i \hat n_j + \sum_j\epsilon_j \hat n_j,
  \label{eq:gEHM}
\end{eqnarray}
where $u_{ij}$ and $v_{ij}$ are independent random variables with values $\pm 1$. 
In the large-disorder limit a (pseudo-)spin resonance analysis~\cite{Burin2006,Yao2014,Gutman2016} shows that the isotropic EHM (without random signs) 
has a critical dimension $d_c = (b+2)/2$, that is, in 1D MBL can appear for $0<b<2$. The EHM with random signs has a critical dimension $d_c = b/2$, that is, MBL 
cannot appear for $b<2$ in the thermodynamic limit.

\begin{figure}[t]
  \begin{center}
     \includegraphics[width =1.0\columnwidth]{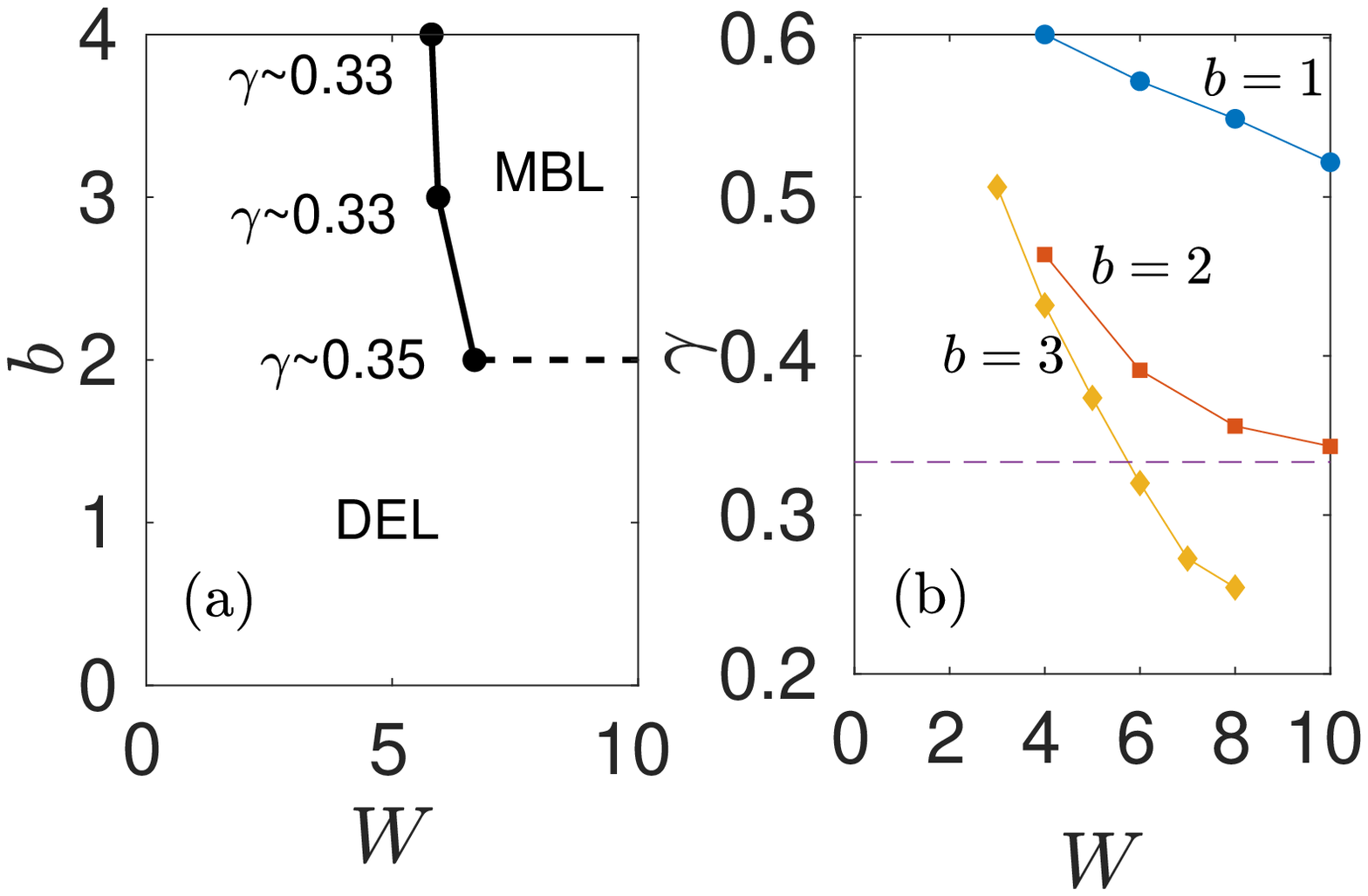}
    \vspace*{-0.6cm}
    \caption{(Color online) (a) Phase diagram of Model~\eqref{eq:gEHM} as a function of the disorder strength $W$ and the interaction power $b$.  
    The MBL transition is obtained by performing a finite-size scaling of our results of $\langle r \rangle$ for various system sites up to $L=18$. 
    For $b<2$ we observe localization for large $W$ although we associate it to the small size of the considered systems.
    (b) Exponent $\gamma$ of the growth of entanglement entropy 
    as a function of $W$ for $b=1$, $2$, and $3$. Note that for $b=2$ the exponent $\gamma$ approaches asymptotically $0.33$ as expected 
    if the critical interaction exponent was $b_c=2$, and the critical growth exponent $\gamma_c=0.33$.}
    \label{fig:SM4}
    \vspace{-0.6cm}
  \end{center}
\end{figure}

Using the same techniques as in the main text, we have obtained the phase diagram and the time evolution of the entanglement entropy for Model~\eqref{eq:gEHM}. 
Our results, depicted in Fig. \ref{fig:SM4} (a), are consistent with the (pseudo)-spin analysis. 
At the MBL transition, the entanglement entropy grows algebraically with exponent $\gamma_c\approx0.33$. 
For $b=2$, $\gamma$ approaches $\gamma_c$ with increasing disorder strength~(Fig.~\ref{fig:SM4} (b)), as one could expect if the critical value of $b$ is $b_c=2$. 
These results, as well as those of the main text, suggest a close relation between the critical $\gamma_c$ of the entanglement entropy growth 
and the critical interaction exponent for observing MBL.
 


